\newcommand{\CLASSINPUTtoptextmargin}{2cm}
\newcommand{\CLASSINPUTbottomtextmargin}{2.4cm}
\documentclass[conference]{IEEEtran}
\IEEEoverridecommandlockouts
\usepackage{comment}
\usepackage{amsmath,amssymb,amsfonts}
\usepackage{algorithmic}
\usepackage{multirow, multicol}
\usepackage{graphicx}
\usepackage{textcomp}
\usepackage{xcolor}
\usepackage{stfloats}
\usepackage[sorting=none]{biblatex} 
\def\BibTeX{{\rm B\kern-.05em{\sc i\kern-.025em b}\kern-.08em
    T\kern-.1667em\lower.7ex\hbox{E}\kern-.125emX}}
\begin{document}

\title{Scalable Wi-Fi RSS-Based Indoor Localization via Automatic Vision-Assisted Calibration}

\author{\IEEEauthorblockN{Abdulkadir Bilge, Erdem Ergen, Burak Soner, Sinem Çöleri}
\IEEEauthorblockA{\textit{Koç University} \\
Istanbul, Türkiye \\
\{abilge20, eergen20, bsoner, scoleri\}@ku.edu.tr}}

\maketitle

\begin{abstract}
Wi-Fi-based positioning promises a scalable and privacy-preserving solution for location-based services in indoor environments such as malls, airports, and campuses. RSS-based methods are widely deployable as RSS data is available on all Wi-Fi-capable devices, but RSS is highly sensitive to multipath, channel variations, and receiver characteristics. While supervised learning methods offer improved robustness, they require large amounts of labeled data, which is often costly to obtain. We introduce a lightweight framework that solves this by automating high-resolution synchronized RSS-location data collection using a short, camera-assisted calibration phase. An overhead camera is calibrated only once with ArUco markers and then tracks a device collecting RSS data from broadcast packets of nearby access points across Wi-Fi channels. The resulting (x, y, RSS) dataset is used to automatically train mobile-deployable localization algorithms, avoiding the privacy concerns of continuous video monitoring. We quantify the accuracy limits of such vision-assisted RSS data collection under key factors such as tracking precision and label synchronization. Using the collected experimental data, we benchmark traditional and supervised learning approaches under varying signal conditions and device types, demonstrating improved accuracy and generalization, validating the utility of the proposed framework for practical use. All code, tools, and datasets are released as open source.

\end{abstract}

\begin{IEEEkeywords}
Wi-Fi positioning, received signal strength (RSS), vision-based calibration, robust localization, deep learning.
\end{IEEEkeywords}

\section{Introduction}

Indoor localization supports many applications in retail, airport navigation, and security, especially when individuals or assets with mobile devices require self-localization \cite{liu2020survey}. Among available methods, Wi-Fi-based positioning using signals from stationary access points (APs) is particularly popular due to its passive nature and the ubiquity of Wi-Fi infrastructure. Unlike camera-based systems, it also preserves privacy since user Wi-Fi devices are uniquely identifiable without personal information, avoiding the trade-offs between spatial resolution and privacy concerns \cite{zhao2018privacy}.

Wi-Fi based localization methods use different characteristics of the Wi-Fi signal: the received signal strength (RSS), the channel state information (CSI), or signal round-trip-time (RTT). CSI and RTT approaches use detailed physical-layer data such as round-trip time \cite{cao2020indoor} and per-subcarrier complex phase-amplitude characteristics \cite{chen2023}, and can achieve even sub-centimeter accuracy in controlled environments \cite{he2024, du2024}. However, their deployment is limited due to hardware constraints: CSI and RTT data are not accessible on most commercial devices without specialized tools such as Nexmon \cite{nexmon} or Intel CSI Tool \cite{intelcsitool}, or custom firmware \cite{ahmad2024}. Among these methods, RSS remains the most practical signal type for localization as it is available on virtually all devices \cite{dai2023}.

\begin{figure}[t]
  \centering
  \includegraphics[width=0.45\textwidth]{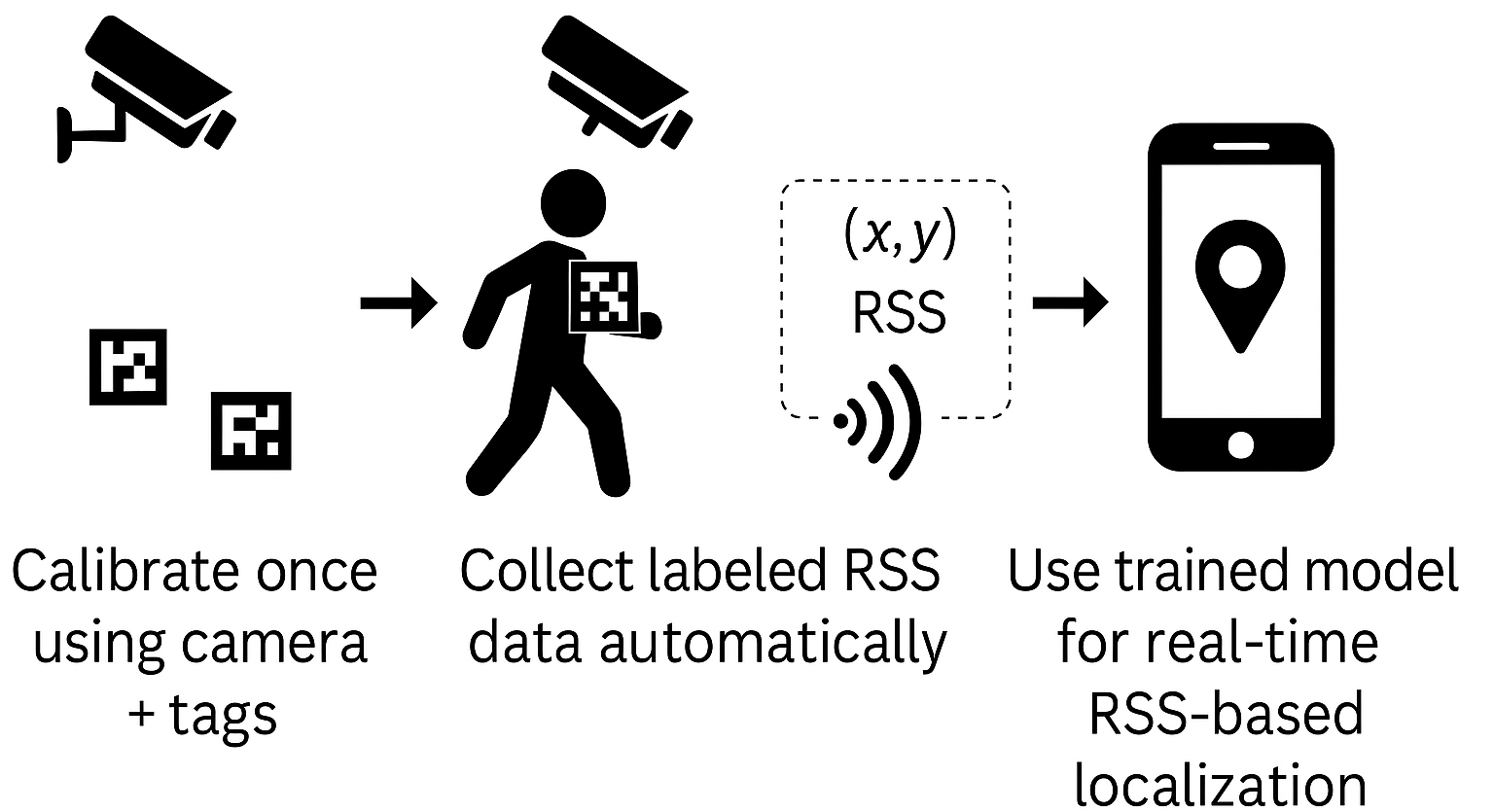}
  \caption{One-time calibration with visual object trackers and markers allows automatically collecting labeled RSS-location data without tedious manual annotation. Ample data allows trained models to generalize with high accuracy, unlocking scalable localization using RSS alone.}
  \label{teaser}
  \vspace{-2mm}
\end{figure}

Despite its practicality, RSS-based localization suffers from significant robustness issues. Traditional fingerprinting and geometric methods (e.g., RSS signal model-based triangulation) are affected by environmental changes, device variability \cite{ahmad2024, dai2023}, and especially multipath effects. Since Wi-Fi systems are designed for coverage rather than directionality, the functional relationship between RSS and location is often weak and unreliable. Dynamic factors such as human motion or layout changes further degrade performance and require frequent recalibration. Data-driven methods using neural networks and large training datasets have been proposed to improve robustness \cite{feng2021}, but they often involve costly data collection and generalize poorly across settings—such as between rooms or across different devices. To address these limitations, recent work has explored auxiliary sensing methods like vision-based tracking for automated calibration. Examples include webcam- or smartphone-based self-labeling \cite{radaelli2014cambasedcalib, kim2024smartphone}, but these lack a complete analysis of how calibration quality impacts localization performance.

In this work, we present a lightweight, vision-assisted RSS based localization framework that automates data collection using a calibrated overhead camera (e.g., a ceiling-mounted security camera). The system uses ArUco markers for initial calibration and tracks a marker-attached device during a short data collection phase, producing synchronized RSS and location pairs at high spatial resolution. Using this setup, we collect experimental datasets and benchmark a range of localization models, including both traditional and supervised learning-based methods. We systematically evaluate how well these models generalize under different conditions, such as varying signal strengths across environments and receiver device types. The system is modular, practical, and open-source, designed for easy deployment and reproducible benchmarking of RSS-based localization algorithms.

\begin{figure*}[b]
  \centering
  \includegraphics[width=0.9\textwidth]{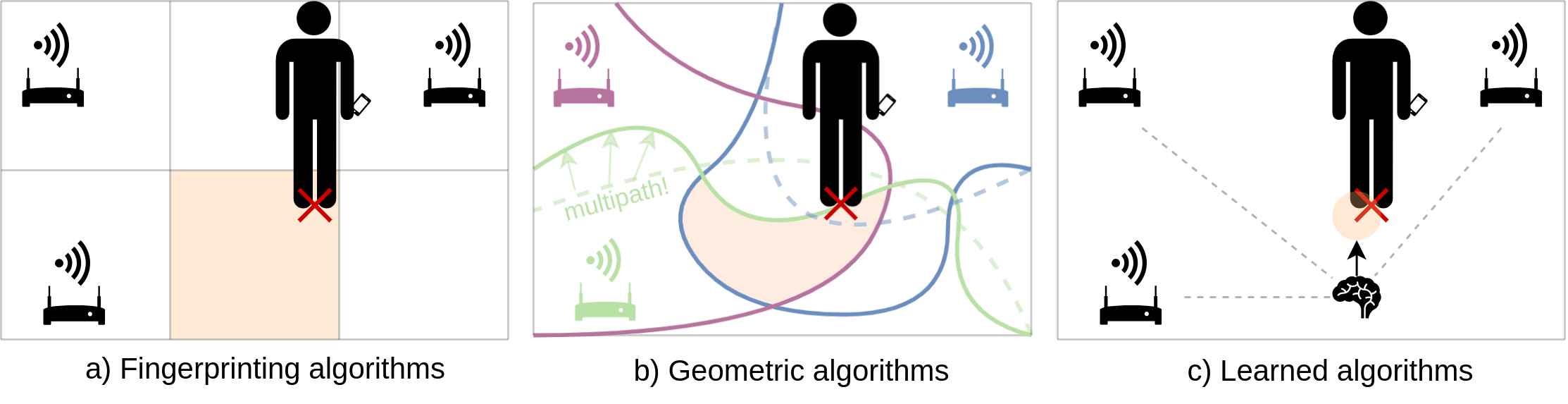}
  \caption{RSS-based localization: (a) fingerprinting methods predict the nearest neighbor, resulting in coarse resolution; (b) geometric methods rely on signal models and are sensitive to multipath; (c) data-driven methods learn robust mappings from RSS to location across devices and environments.}
  \label{rss_comparison}
\end{figure*}

\section{Related Work}

Two main families of traditional RSS-based indoor localization methods are fingerprinting and geometric approaches. Fingerprinting relies on an offline phase where RSS values are collected at known locations and an online phase where real-time measurements are matched to this database. A common technique is nearest neighbor classification in the raw RSS feature space. While this approach can offer robustness when the data is rich, it suffers from coarse spatial resolution, as dense fingerprinting requires labor-intensive site surveys.

In contrast, geometric methods estimate the location of a device by modeling signal propagation directly (e.g., finding position fixes at the intersections of modeled "equi-RSS" hyperbola curves). These methods can theoretically provide fine-grained localization, but perform poorly in Wi-Fi environments, where access points are designed for omnidirectional coverage and heavily exploit multipath. Consequently, assumptions about equi-distance signal contours break down under multipath and non-line-of-sight conditions. Hybrid approaches that interpolate between sparse fingerprint locations using signal propagation models have also been proposed, but these are difficult to scale across environments due to variability in signal characteristics. 

Recent supervised learning–based methods aim to address this tradeoff between robustness and resolution, as depicted in Fig. \ref{rss_comparison}. One class of approaches still performs classification, but in a learned feature space rather than on raw RSS \cite{giney2020wi}. Another class performs regression from RSS measurements to physical coordinates. These models, which also recently explore using modern architectures such as Transformers \cite{wu2024attention} and prototypical networks \cite{ma2023prototypicalnets}, can achieve higher spatial resolution, but require large amounts of labeled data, which is expensive to collect. Our proposed framework directly addresses this challenge by automating the collection of high-resolution labeled data via vision-assisted calibration.

\begin{figure}[t]
  \centering
  \includegraphics[width=0.90\linewidth]{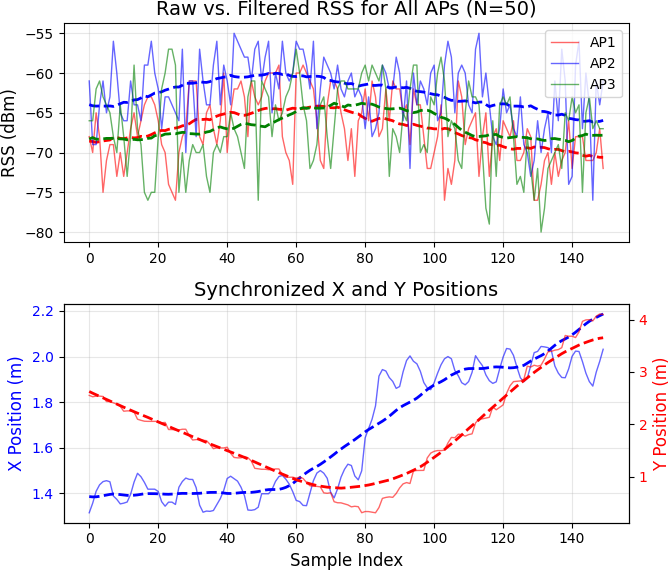}
  \caption{Filtering collected data for higher correlation between RSS (top) and location ($x,y$) data (bottom). Dashed lines indicate filtered versions.}
  \label{fig:filtering}
\end{figure}

\begin{figure}[t]
  \centering
  \includegraphics[width=0.46\textwidth]{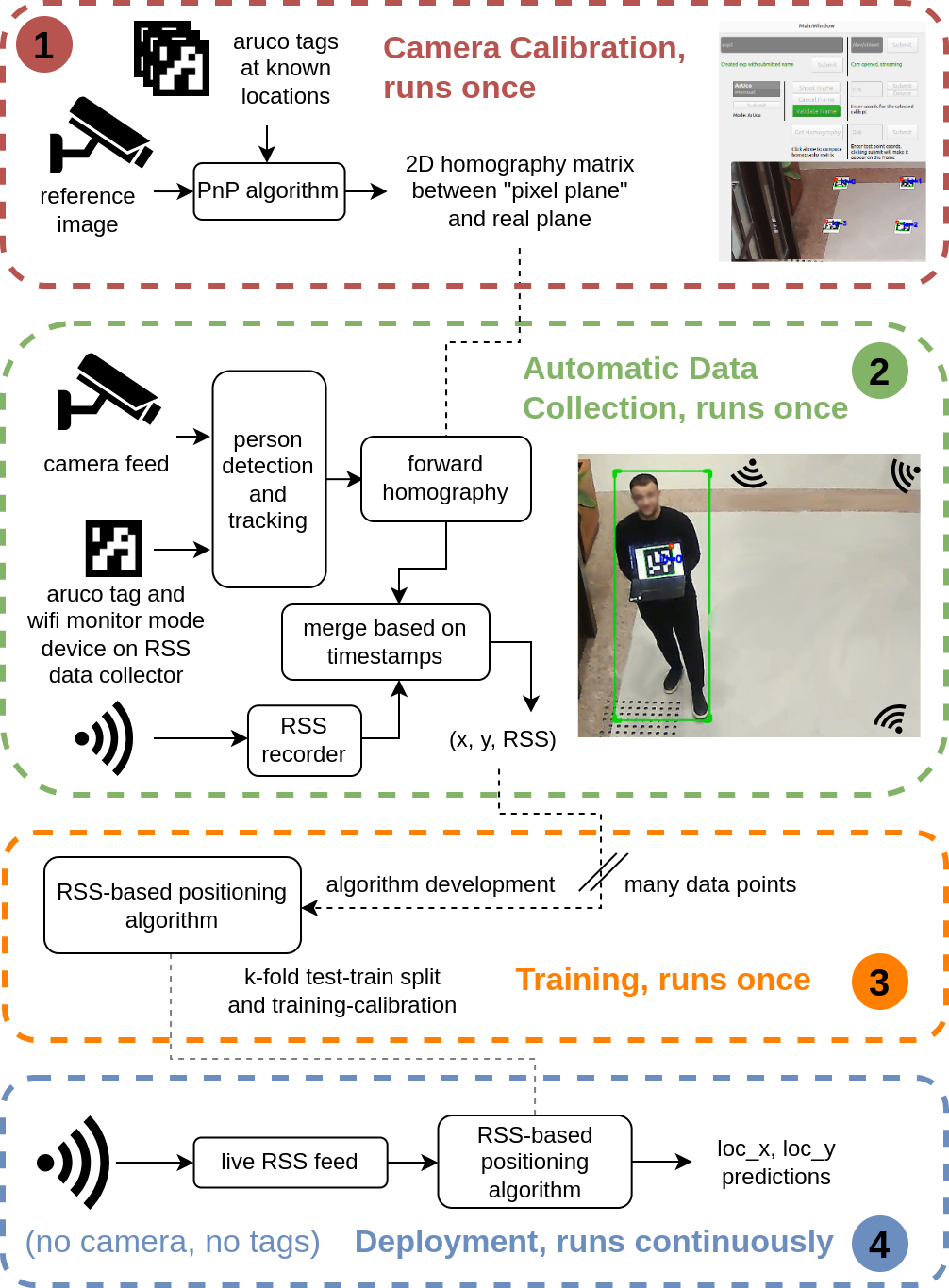}
  \caption{System architecture of the proposed framework. The process consists of four stages: (1) one-time camera calibration, (2) automatic data collection with synchronized RSS and visual tracking, (3) algorithm training, and (4) continuous real-time localization using only RSS input.}
  \label{system}
\end{figure}

\section{Proposed Framework}

Our framework automates high-resolution RSS–location data collection and end-to-end model training using a short, vision-assisted calibration. The setup involves two synchronized components: (i) an overhead camera for ground-truth location labeling, and (ii) an RSS collector operating in Wi-Fi monitor (a.k.a. promiscuous) mode. The vision pipeline is discarded after the calibration phase, and only the RSS data on end-user devices are used for positioning.

The camera is calibrated once, using ArUco markers placed at known positions, with the standard Perspective-n-Point (PnP) algorithm \cite{li2012pnp}. During data collection, the (x,y) location of the person carrying the RSS collector is tracked using a YOLOv8+SORT pipeline (carries an ArUco marker for identification), and the RSS collector captures broadcast Wi-Fi packets in monitor mode at 10 Hz (AP broadcast messages), channel hopping across Wi-Fi bands to capture rich multipath characteristics. 

We apply a moving average filter with a fixed-size window to both RSS and (x,y) streams, as shown in Fig. \ref{fig:filtering}, to improve correlation between RSS and location data (e.g., correlation coefficients are 2x higher for a window of 50 samples). However, this also bounds the user to slow movements (both during calibration and deployment) since with a sampling rate of 10 Hz for RSS, a window of e.g., $N{=}50$ samples means 5 seconds of movement is averaged. While much smaller windows can also be used for capturing faster movements, this slow-sampling problem of RSS-based position estimation due to slowly repeating broadcast messages is well-known \cite{liu2020survey}. 

After data acquisition, the framework trains localization models using the collected data. These include traditional nearest-neighbor classifiers, signal model-based interpolation approaches, and supervised learning models. Since the filtering step described above improves correlation, using convolutional neural networks (CNN), which are basically nonlinear combinations of such FIR filters, is a sensible choice while benchmarking supervised learning methods: A compact 1-D CNN over fixed-length windows captures these local temporal dependencies with a small number of parameters, making them easier to deploy on mobile platforms compared to other sophisticated architectures such as LSTMs and Transformers. Moreover, since CNNs are stationary models in time (i.e., inference at each time step is independent of inference outputs in other timesteps, the model has no memory), we can randomize the few training samples we have and remove the bias of specific motion profiles, achieving better generalization. Trained models are then exported for inference on end-user mobile devices, which require only RSS input for real-time localization—no camera, markers, or additional infrastructure.

The framework, depicted in Fig. \ref{system}, includes user-friendly graphical interfaces for calibration, data collection and training. This enables rapid and automated deployment of robust localization models for any environment, such as against varying crowd densities and different receiver devices.

\section{Accuracy Analysis}
We quantify the accuracy limits of our vision-assisted calibration pipeline, which bound the performance of any RSS-based model trained on its labels. The analysis covers two aspects: (i) \emph{spatial} errors from pixel-to-world mapping and (ii) \emph{temporal} errors from synchronizing RSS with visual labels.

\subsection{Spatial Attributes}\label{sec:spatial}

Let $\sigma_t$ be the standard deviation of the estimation error between the true and estimated 2-D position on the floor plane (i.e., perpendicular projection). We model the total spatial error variance\footnote{The effect of device height and other minor effects like condition number of the homography matrix are ignored here since they are negligible compared to the other terms considered for the proposed setup.} as the sum of four independent components:

\begin{equation}
  \sigma_t^2
  \;=\;
  \sigma_{\text{px}}^2
  \;+\;
  \sigma_{\text{tag}}^2
  \;+\;
  \sigma_{\text{det}}^2
  \;+\;
  \sigma_{\text{foot}}^2 ,
  \label{eq:sigma_t}
\end{equation}

\noindent where the individual terms are defined as follows:

\vspace{2mm}
\begin{itemize}
    \item \emph{Pixel–scale quantization \(\sigma_{\text{px}}\)} captures the limit imposed by finite sensor resolution; each pixel subtends a fixed ground-plane distance determined by camera height and field of view.
    \vspace{2mm}
    \item \emph{Tag-placement error \(\sigma_{\text{tag}}\)} arises because ArUco markers cannot be positioned exactly at their nominal locations.  The resulting perturbation propagates through the PnP solution and affects every reconstructed point.
    \vspace{2mm}
    \item \emph{Detector–tracker variance \(\sigma_{\text{det}}\)} reflects frame-to-frame jitter of the YOLOv8 detector and SORT tracker, measured empirically from annotated video frames (sampling in our setup showed this to be $\approx$ 3 px in most scenarios).
    \vspace{2mm}
    \item \emph{Foot-point/ground-plane heuristic \(\sigma_{\text{foot}}\)}: The framework projects the lower midpoint of the bounding box onto the floor plane and assumes that is the device position. Camera angles and user movements induces a systematic error in the location label.
    \vspace{2mm}
\end{itemize}

\vspace{2mm}

The four variance terms in~\eqref{eq:sigma_t} were simulated (10k Monte-Carlo iterations) for different settings to characterize these spatial error attributes. We provide the source code and details for this simulator in our open source repository. Table~\ref{tab:spatial_error} summarizes the results for our \emph{baseline} installation (3m camera height, 5m vertical field of view, tag misplacement $\sigma_{\text{tag}}=$0.10m) and for two enlarged-FoV scenarios that illustrate how localization accuracy degrades with both camera scale and careless marker placement.

For the baseline installation the combined spatial label uncertainty is \(\sigma_{t}\approx 9.7\) cm, well below the meter-level RMSE reported by state-of-the-art RSS localization systems \cite{feng2021}. For other configurations, such as when the camera FoV is increased, the homography-related sensitivities dominate, raising \(\sigma_{t}\) to 27 cm.  If, in addition, marker placement errors are increased to 30 cm variance, the tag-placement term becomes the principal contributor and the overall uncertainty rises to 76 cm.  These results confirm that moderate tag placement errors around 5-10 cm do not cause critical label uncertainty levels, but large placement errors and low pixel-density camera FoV do cause the label errors to go up to the meter level.

\subsection{Temporal Attributes}\label{sec:temporal}

Let $\Delta t$ be the time misalignment between a camera frame and its nearest RSS sample (in time). If the device moves at speed $v$, the induced spatial error is $\varepsilon_{\text{temp}} = v\,\Delta t$. We can model $\Delta t$ as the maximum of the three dominant latencies
\[
    \Delta t = \max\!\bigl\{1/f_{\text{cam}},\,1/f_{\text{rss}},\,\Delta t_{\text{align}}\bigr\},
\]
where $f_{\text{cam}}\!=\!30$ Hz, $f_{\text{rss}}\!\approx\!10$ Hz (broadcast messages) and $\Delta t_{\text{align}}$ is the clock alignment error between the RSS collector and the vision-assistant collecting the location data. $\Delta t_{\text{align}}$ is bounded to less than 10 ms by using a Network Time Protocol (NTP) provider to match the clocks of the two devices. Then, to bound this timing-induced error, $\varepsilon_{\text{temp}}$, to less than 5 cm,  we need to impose $v\le0.5$ m/s during calibration. A similar latency budget applies at deployment, since methods will need to aggregate at least 5-10 RSS samples before updating the user position, necessitating users to remain static for at least a few seconds \cite{liu2020survey}. 

\subsection{Combined Label Uncertainty}

Combining~(\ref{eq:sigma_t}) with $\varepsilon_{\text{temp}}$ gives the overall 1-$\sigma$ label uncertainty for the data the proposed setup produces:
\[
  \sigma_{\text{label}}
  \;=\;
  \sqrt{\sigma_t^{2} + \varepsilon_{\text{temp}}^{2}}~~,
\]
which serves as an upper bound on the achievable localization error for any algorithm trained using these labels. As shown in the simulations above, for the base configuration we considered, this error is smaller than 10 cm.

\begin{table}[t]
  \centering
  \caption{Simulated label uncertainty due to spatial attributes}
  \label{tab:spatial_error}
  \begin{tabular}{@{}lccccc@{}}
    Scenario & $\sigma_{\text{px}}$ & $\sigma_{\text{tag}}$ &
    $\sigma_{\text{det}}$ & $\sigma_{\text{foot}}$ & $\sigma_{t}$ \\
             &  [m] &  [m] & [m] &  [m] &  [m] \\
    Base (5 m\,FoV, 10 cm~tags)           & 0.0023 & 0.050 & 0.014 & 0.082 & \textbf{0.097} \\
    Large FoV (25 m, 10 cm~tags)              & 0.0116 & 0.251 & 0.069 & 0.082 & \textbf{0.273} \\
    Large FoV (25 m, 30 cm~tags)              & 0.0116 & 0.755 & 0.069 & 0.082 & \textbf{0.762} \\
  \end{tabular}
\end{table}

\section{Experimental Results}

Four independent recordings were collected with the "Base" camera configuration from Table \ref{tab:spatial_error} and three commercial off-the-shelf routers configured as APs, positioned around an approximately rectangular region of 4m x 6m, centered at the camera FoV. The recordings are called Experiments \#5-\#8, where ~\#5–\#7 (\(883, 1865, 2317\) labeled samples) were captured with receiver~A (Intel - iwlwifi), while Exp.~\#8 (\(1092\) samples) used receiver~B (MediaTek - mt7921e), introducing receiver and OS driver variations to the dataset alongside the RSS collector and bystander movement variations present in each recording. We first use this dataset to analyze our framework and the RSS-based localization problem by training simple CNNs \footnote{The CNN: 4 layers with bias and ReLU, kernel sizes are \{13, 11, 9, 7\} and channel sizes are \{8, 16, 8, 2\}; final layer produces $x$ and $y$ predictions.} in different evaluation runs, and then we compare the CNN with two traditional methods to provide a benchmark\footnote{All data, scripts, and trained models are available in our public repository: \url{https://github.com/sonebu/wifi_rss_positioning_visioncalib}}.

\vspace{2mm}
\begin{enumerate}
\item \textbf{kNN}: nearest-neighbor fingerprinting on raw RSS,
\vspace{2mm}
\item \textbf{kNN+Interpolation}: weighted sum of M closest kNN estimations, weights are Euclidean distance to the input RSS vector, thereby assuming a $1/r^2$ signal model where $r$ is distance from the AP.
\vspace{2mm}
\item \textbf{CNN}: the lightweight 1-D convolutional network.
\vspace{2mm}
\end{enumerate}
\vspace{2mm}

\noindent We report three evaluation runs (R\#):

\vspace{1mm}
\begin{itemize}
  \item \textbf{R1) Data–efficiency:} The CNN is trained on random 5\% – 95\% train-test splits of the whole dataset (all 4 recordings) to understand how much data collection will be necessary during calibration for proper generalization. Results show that only a few minutes of calibration per configuration gives the best possible result; an error under the label uncertainty level.
  \vspace{1mm}
  \item \textbf{R2) Generalization:} The CNN is trained on 3 out of 4 recordings, and is tested on the 4th, cycled through all combinations. Results show that the CNN does generalize well to the $y$ axis in similar settings (e.g., Exp. 5-6-7), but fails to provide accurate estimations for the $x$ axis.
  \vspace{1mm}
  \item \textbf{R3) Method comparison:} kNN, kNN+Interp and CNN are compared on the total dataset, where the CNN is the clear overall winner.
\end{itemize}
\vspace{1mm}

\begin{figure}[!t]
  \centering
  \includegraphics[width=\linewidth]{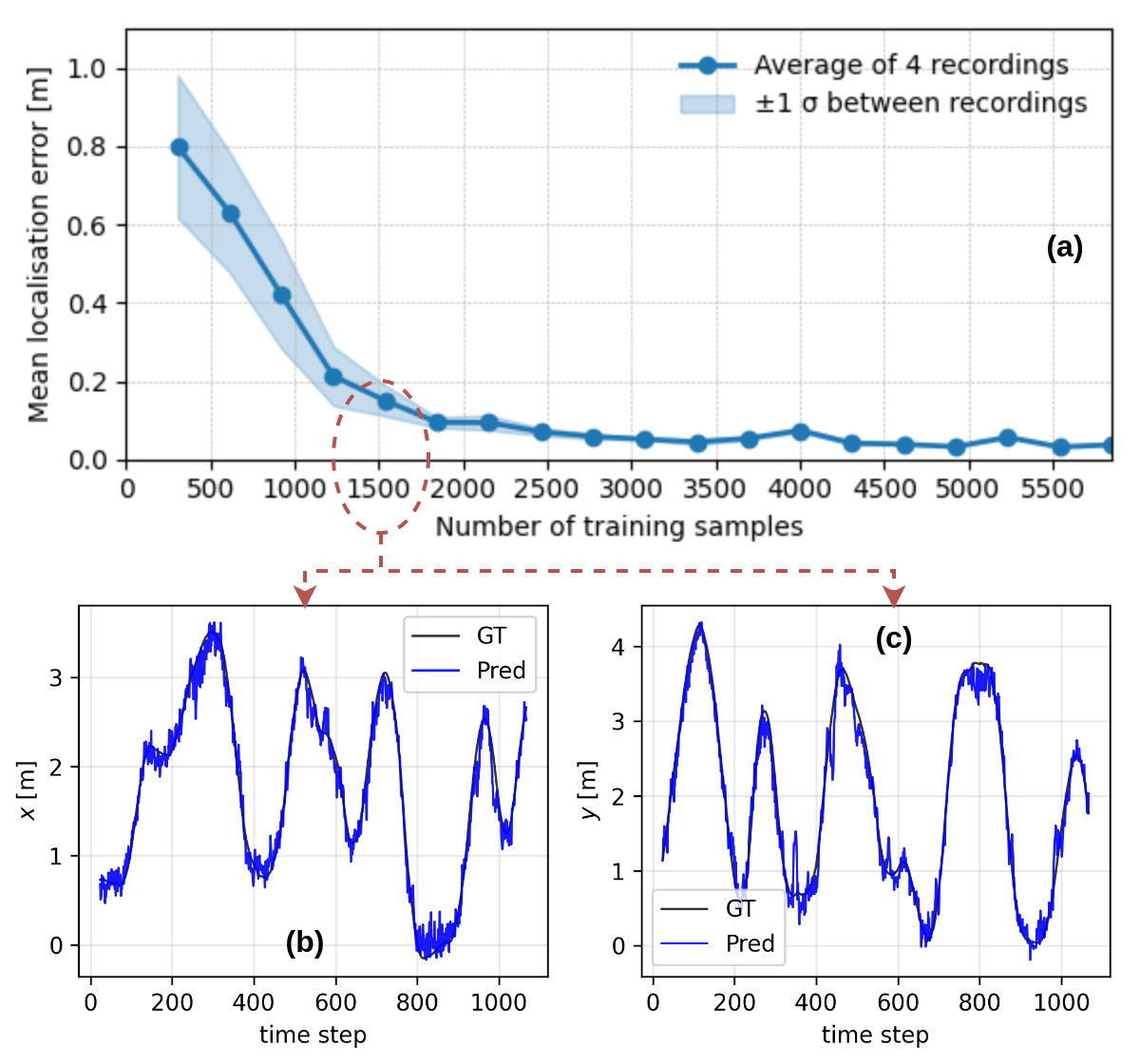}
  \caption{Results in (a) show CNN performance vs. number of labeled samples, averaged over all 4 recordings. Results in (b) and (c) show the $x$ and $y$ estimation performance of the CNN trained with just 1500 samples, on Experiment \#8, the case with receiver B.}
  \label{fig:data_curve}
\end{figure}

\subsection{R1: Data–efficiency}

We merge the 4 recordings and then train-test the CNN on random 5\% – 95\% train-test splits of the dataset. Figure~\ref{fig:data_curve} shows the mean error on the test set with the curve and the standard deviation with the shaded error region, averaged over the 4 experiments versus the fraction of the train-test split. Results show that with only $\approx$150 seconds of calibration (1500 labels) the CNN already attains a result very close to the best possible performance. Note that the label uncertainty for this configuration is around 10 cm, so after $\approx$1500 samples of calibration the error would be dominated by the label uncertainty.

\begin{figure}[t]
  \centering
  \includegraphics[width=0.90\linewidth]{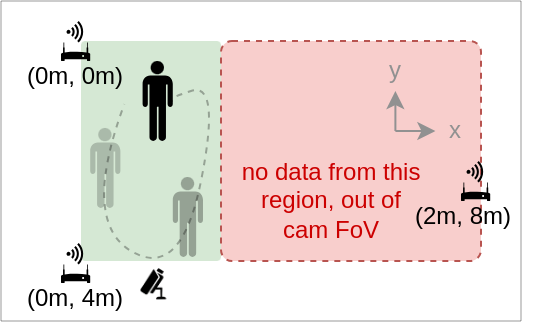}
  \caption{AP placement and RSS collector movement profile during calibration in the experimental setup. Limited coverage in the x axis decreased modeling performance.}
  \label{expsetup}
\end{figure}

\subsection{R2: Cross-recording generalization}

\begin{figure}[t]
  \centering
  \includegraphics[width=\linewidth]{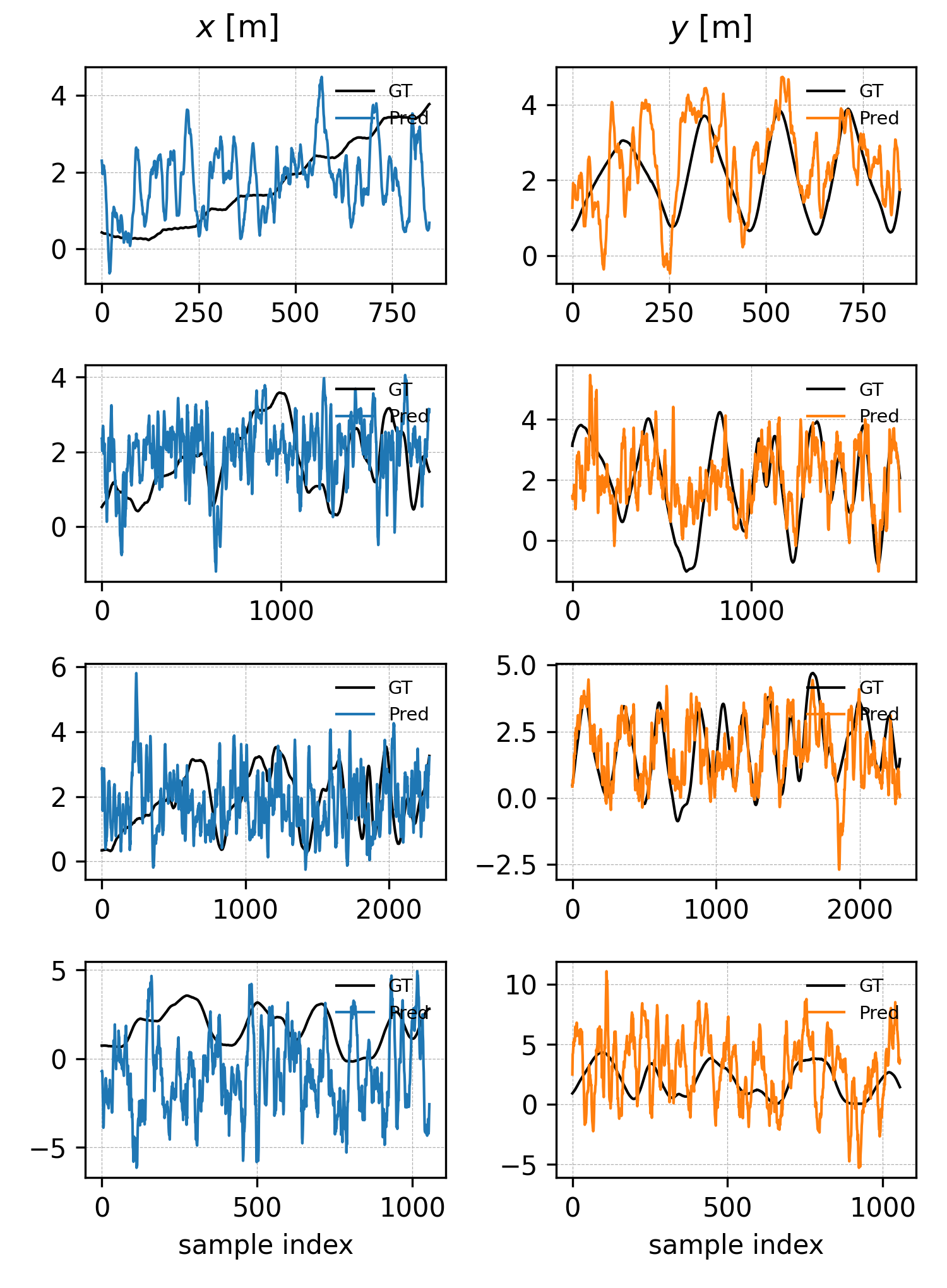}
  \caption{Generalization test: The CNN is trained on 3 out of 4 recordings and evaluated on the 4th (rows show different hold-out combinations, Experiments \#5 to \#8, respectively). Left column shows the performance in $x$ and right column shows the performance in $y$.}
  \label{fig:loro_cnn}
\end{figure}

We train the CNN on 3 out of 4 recordings and test it on the 4th, for all combinations, to show the generalization performance of the data collected and the CNN. The results in Figure \ref{fig:loro_cnn} show that although overall generalization accuracy across unseen domains is not as high as the in-domain case shown in Fig. \ref{fig:data_curve}, it is sufficient for sub-1m estimation, and it is much better for the $y$ axis compared to the $x$ axis. This stems from asymmetric nature of the AP placement in the experimental setup versus the movement profile chosen for calibration. As shown in Figure \ref{expsetup}, while the RSS collector moves over the whole $y$ range during calibration, the $x$ range is not covered as much because of limited camera FoV, resulting in lower sensitivity of the RSS data towards x movements. This highlights the importance of choosing movement profiles during calibration; RSS collectors should cover the whole region defined by the APs in order to produce proper calibration data. Moreover, the effect of receiver variation is also observed; Generalization to Exp. \#8 is not as good as in the other Experiments since there are fewer samples recorded with receiver B, highlighting the importance of using balanced datasets with a few different receiver devices during calibration. 

\subsection{R3: Benchmark, Traditional vs. Learning-based Methods}

We compare the lightweight CNN with two classical RSS-fingerprinting baselines on the full dataset. Table~\ref{tab:method_comp} reports the mean and standard deviation of the $L_{2}$ error on each recording. Plain $k$NN attains $\sim\!2\,$m median accuracy, which improves only marginally when the inverse-square interpolation model is added. In contrast, the CNN pushes the error below \mbox{15 cm} in all cases (\(<10\) cm on receiver~A). These results confirm that once high-quality labels are available, such as with the proposed data collection and calibration framework, even a lightweight network can deliver cm-scale positioning performance with substantially better generalization than traditional methods in the literature.

\begin{table}[t]
  \centering
  \caption{Traditional vs learning-based methods, mean ± std. deviation.}
  \label{tab:method_comp}
  \begin{tabular}{|l|c|c|c|}
    \hline
    ID & kNN & kNN+Interp & \textbf{CNN (ours)} \\
    \hline
    Exp-5 & 1.884 ± 0.981 m  & 1.597 ± 0.793 m  & \textbf{0.094 ± 0.093 m } \\
    Exp-6 & 1.874 ± 1.103 m  & 1.585 ± 0.885 m  & \textbf{0.097 ± 0.093 m } \\
    Exp-7 & 1.865 ± 1.003 m  & 1.584 ± 0.838 m  & \textbf{0.097 ± 0.090 m } \\
    Exp-8 & 2.376 ± 1.074 m  & 2.037 ± 0.961 m  & \textbf{0.144 ± 0.150 m } \\
    \hline
  \end{tabular}
\end{table}

\subsection{Limitations}

We recognize the following limitations in this study: Our evaluation spans a single site with three AP layouts and two different receivers, but broader datasets (device diversity, camera FoV/lighting, crowd density, and edge-case motion profiles) would enhance applicability. We benchmark a compact 1-D CNN for deployability with limited data, but richer sequence/attention models and unsupervised/self-supervised adaptation may further improve cross-device/site generalization when more data/compute are available. Moreover, long-term drift and AP churn were not studied, but lightweight online adaptation or periodic re-calibration approaches can be investigated. These are left to future work.

\section{Conclusion}

We presented a lightweight framework that combines a one-time, vision-assisted calibration with monitor-mode RSS collection to generate cm-accurate location labels at scale for Wi-Fi RSS-based positioning. Fundamental analyses showed that the spatial–temporal uncertainty introduced by the camera, marker placement and label synchronization remains below 10 cm. Using the collected labels, we trained a compact CNN and demonstrated that it generalizes much more accurately with high-quality data compared to existing methods (e.g., $k$NN + $1/r^2$ interpolation) for unseen recordings and different receivers. These results confirm that high-quality labels provided by our lightweight framework are key for robust RSS-based Wi-Fi positioning. Future work will explore self-supervised adaptation and Transformer-based modeling for automatic adaptation to new environments, potentially without requiring vision-based calibration.

\printbibliography

\end{document}